\documentclass{aastex}
\usepackage{spr-astr-addons}
\begin{document}
\title{Cosmological parallax--distance formula}
\shorttitle{Cosmological parallax--distance formula}
\shortauthors{Singal}

\author{Ashok K. Singal\altaffilmark{1}}

\altaffiltext{1}{Astronomy and Astrophysics Division, Physical Research Laboratory,
Navrangpura, Ahmedabad - 380 009, India. Email:asingal@prl.res.in}
\abstract{
The standard cosmological parallax--distance formula, as found in the literature, 
including text-books and reference books on cosmology, requires a correction. This correction stems from the 
fact that in the standard text-book derivation it has been ignored that any chosen baseline in a gravitationally bound 
system does not partake in the cosmological expansion. Though the correction is available 
in the literature for some time, the text-books still continue to use the older, incorrect 
formula, and its full implications are not yet fully realized. 
Apart from providing an alternate correct, closed-form expression that is more suitable and convenient 
for computations for certain limiting cases of FRW ($\Lambda=0$) world models, 
we also demonstrate how one can compute parallax distance for the currently favored flat-space accelerating-universe 
($\Lambda>0, k=0$) cosmologies. Further, we show that the correction in parallax distance 
at large redshifts could amount to a factor of three or even more. Moreover, even in an 
infinite universe the parallax distance does not increase indefinitely with redshift  
and that even the farthest possible observable point may have a finite parallax angle, a factor that needs to be carefully 
taken into account when using distant objects as the background field against which the parallax of a 
foreground object is to be measured. Some other complications that could arise in parallax measurements of a distant 
source, like that due to the deflection of incoming light by the gravitation field of the Sun and other planetary bodies in 
the solar system, are pointed out. 
}
\maketitle
\section{\bf INTRODUCTION}
Cosmological distance formulae are standard formulae for different world models in use for more 
than half a century and are available in review articles and text books 
on cosmology. The two most commonly used formulae for cosmological distant sources are the 
luminosity-distance formula and the angular diameter-distance formula. Their usage depends upon a sequence of 
calibrations along the cosmic distance ladder (Weinberg 1972) and any systematic over or underestimates in the intermediate 
steps would affect the final distance estimates. Additionally, all errors in the various steps along the distance 
ladder could get accumulated. Moreover, a use of either of these in conjunction with the measured values of flux 
densities or angular sizes of a suitable sample, 
for ascertaining the geometry of the universe, depends upon the assumption of the existence of a standard 
candle/rod, and a successful interpretation is very much dependent on the cosmological evolution of 
the intrinsic luminosities/physical sizes of the parent population of the sources of interest. 
Not only the large spread in their intrinsic values makes it very uncertain to define a standard 
candle/rod, in fact in most cases the effects of the evolutionary changes in source properties are 
overwhelmingly larger than those expected due to differences in geometry between different world models. 
The other two formulae, which presently are being used only for nearby objects, are the proper motion--distance 
formula and the parallax--distance formula. 

For testing different world models, application of the parallax distance in particular, unlike other 
distance measures, does not depend upon any assumption about the intrinsic properties of the observed 
sources. Thus it is also independent of the chosen frequency band since no source property is involved, 
as long as the source is detectable in that band. Moreover, it does not depend upon any intermediate steps with errors 
or possibility of an outright miscalculation along the way. At a given redshift, the observed parallax depends 
only on the chosen baseline of the observer and the adopted world--model geometry. Thus all one requires is a suitable baseline 
for parallax measurements. And as parallax angle is directly proportional to the length of the baseline, the fractional error in 
angle measurements is directly proportional to the fractional errors in  the baseline determination and which could be 
very small. Weinberg (1970) showed that even otherwise, the measurement of redshift and luminosity 
(or angular diameter) distance cannot in principle determine the sign and magnitude of the spatial curvature unless supplemented 
with a dynamical model. However, this ambiguity can be resolved by parallax measurements at cosmological 
distances. Consequently the parallax distance could play an important role in cosmology (Rosquist 1988; R\"as\"anen 2014). 
With the achieved angular resolutions already in the micro-arcsec domain (Fomalont \& Kobayashi 2006),
it could with the advancement of technology, become important rather sooner than expected.

The cosmological parallax--distance formulation is available in text-books on cosmology (Weinberg 1972; Peacock 1999), 
reference books (Lang 1980), 
or review articles (von Hoerner 1974) with expressions for the different world models 
given there. There is a subtle correction required in the standard formulation found in the above references.  
The correction stems from the fact that the physical dimensions of a gravitationally 
bound system like that of the solar system (or for that matter even of larger systems like that of a galaxy), do not 
change with the cosmological expansion of the universe. Therefore two ends of a baseline, used by the observer for 
parallax measurements, do not partake in the free-fall-like motion of the cosmic fluid and thus cannot be considered 
to form a set of comoving coordinates, contrary to what seems to have been implicitly assumed in the standard 
text-book derivations of the parallax--distance formula.  The required correction has been pointed out in a number of 
papers (Novikov 1977; Kardash\"ev 1986; Rosquist 1988; R\"as\"anen 2014) showing how to make the correction needed 
in the parallax formula for the Friedmann-Robertson-Walker (FRW) world models.

That the correction could be substantial (specially for large cosmological distances) can be seen from the following. 
Let us take the simple case of a Euclidean geometry. A baseline of length $b$ will subtend an angle $\Psi$ at a distance 
$d$ in a direction normal to the baseline as,
\begin{equation}
\label{eq:20}
\Psi \approx \frac{b}{d}
\end{equation}
Let us introduce expansion of the universe. 
Because the baseline does not expand along with the underlying cosmic substratum, the two comoving coordinates at the 
locations of the baseline ends will have a relative expansion speed $v=H_{\rm o} b$ with respect to the rigid baseline. Due to
this, two observers comoving with the substratum, because of the relative motion, will have a stellar aberration, 
$\delta \approx b H_{\rm o}/c$. This will reduce the value of the parallax angle calculated otherwise and then the  
parallax angle inferred by the comoving observers will be, 
 
\begin{equation}
\label{eq:21}
\psi \approx  \frac{b}{d}-\frac{bH_{\rm o}}{c}=\frac{b}{d} \left(1-\frac{H_{\rm o}d}{c}\right) 
= \Psi \left(1-\frac{H_{\rm o}d}{c}\right) \\
\end{equation}

As we see that the factor $H_{\rm o}d/c$ within the parentheses could be substantial for $d$ that is of the order of $c/H_{\rm o}$, 
(cosmological distances!), and the parallax angle could be only a fraction of what calculated otherwise.

Here we shall derive the parallax distance in an alternate, closed form (as a function of 
redshift) for FRW world models, for $\Lambda=0$ cosmologies. This alternate form is especially convenient for computing 
the parallax distance in the limiting case of the deceleration parameter and/or redshift. We shall also show how the parallax 
distance can be calculated for the more modern $\Omega_\Lambda \neq 0$ cosmologies, demonstrating for some representing cases. 
First we shall in the next section briefly review the formulation given in the literature and in the 
section following that, we spell out the required correction. As we will show 
the correction can be very large (a factor of three or even more) at large redshifts.

\section{FORMULATION}
In a homogeneous and isotropic universe, the line element can be expressed in the Robertson-Walker metric form 
(Weinberg 1972; Peacock 1999),
\begin{eqnarray}
\label{eq:1}
{\rm d}s^2=c^2 {\rm d}t^2 - R^2(t)\left[\frac{{\rm d}r^2}{\left(1-k\,r^2\right)^{1/2}}+
r^2({\rm d}\theta ^2+\sin^2\theta\; {\rm d}\phi^2)\right],
\end{eqnarray}
where $R(t)$, a function of time $t$, is known as the cosmic scale factor, $k$ is the curvature index that can take one 
of the three possible values $+1, 0$ or $-1$ and $(r, \theta, \phi)$ are the time-independent comoving coordinates.

From Einstein's field equations, one can relate the curvature index $k$ and the present values of the cosmic 
scale factor $R_{\rm o}$ to the Hubble constant $H_{\rm o}$, the matter energy density $\Omega_m$ and the vacuum energy (dark energy) 
density $\Omega_\Lambda$ as (Peacock 1999),
\begin{eqnarray}
\label{eq:2}
\frac{k\,c^2}{H_{\rm o}^2R^2_{\rm o}}=\Omega_m+\Omega_\Lambda-1.
\end{eqnarray}
The space is flat ($k=0$) if $\Omega_m+\Omega_\Lambda=1$. 

As shown by Weinberg (1972), an observer using a baseline $b$ to make measurements of a source at radial 
coordinate $r$, in a direction normal to the baseline, will infer a parallax angle, 
\begin{eqnarray}
\label{eq:3}
\psi=\frac {b\left(1-k\,r^2\right)^{1/2}} {R_{\rm o}\,r}.
\end{eqnarray}
Defining a parallax distance, as in Euclidean geometry, by $d_{\rm p}=b/\psi$ we can write,
\begin{eqnarray}
\label{eq:4}
d_{\rm p}=\frac{R_{\rm o}\,r}{\left(1-k\,r^2\right)^{1/2}}.
\end{eqnarray}
In general it is not possible to express $d_{\rm p}$ in terms of the cosmological redshift $z$ of the source 
in a close-form analytical expression and one may have to evaluate it numerically. For example, in the   
$\Omega_m+\Omega_\Lambda=1, \Omega_\Lambda \neq 0$ world-models, $r$ is given by (Peacock 1999),
\begin{eqnarray}
\label{eq:5}
r=\frac{c}{H_{\rm o}R_{\rm o}}\int^{1+z}_{1}\frac{{\rm d}z}{\left(\Omega_\Lambda+\Omega_mz^3\right)^{1/2}},
\end{eqnarray}
and since $k=0$, from Eq.~(\ref {eq:4}) one can write,
\begin{eqnarray}
\label{eq:6}
d_{\rm p}=R_{\rm o}\,r=\frac{c}{H_{\rm o}}\int^{1+z}_{1}\frac{{\rm d}z}{\left(\Omega_\Lambda+\Omega_mz^3\right)^{1/2}}.
\end{eqnarray}
For a given $\Omega_\Lambda$, one can evaluate $d_{\rm p}$ from Eq.~(\ref{eq:6}) by a numerical integration. 

However for $\Omega_\Lambda = 0$ cosmologies, where the deceleration parameter $q_{\rm o}=\Omega_m/2$, it is 
possible to express $r$ in an analytical form (Mattig 1958), 
\begin{eqnarray}
\label{eq:7}
r=\frac{c}{H_{\rm o}R_{\rm o}}\frac{q_{\rm o}z+\left(q_{\rm o}-1\right)\left(-1+\sqrt{1+2q_{\rm o}z}\right)}{q_{\rm o}^2(1+z)}.
\end{eqnarray}
Also Eqs.~(\ref{eq:2}) now becomes,
\begin{eqnarray}
\label{eq:8}
\frac{k\,c^2}{H_{\rm o}^2R^2_{\rm o}}=2q_{\rm o}-1.
\end{eqnarray}
Then substituting Eqs.~(\ref{eq:7}) and (\ref{eq:8}) in Eq.~(\ref{eq:4}), it is straightforward to get,
\begin{equation}
\label{eq:9}
d_{\rm p}=\frac{c}{H_{\rm o}}\frac{q_{\rm o}z+\left(q_{\rm o}-1\right)\left(-1+\sqrt{1+2q_{\rm o}z}\right)}
{\left[q_{\rm o}^4\left(1+z\right)^2- \left(2q_{\rm o}-1\right)\left\{q_{\rm o}z + \left(q_{\rm o}-1\right)\left(-1 +
\sqrt{1+2q_{\rm o}z}\right)\right\}^2\right]^{1/2}}.
\end{equation}

This is the general expression for the cosmological parallax distance available in the literature (Weinberg 1972; Lang 1980), and is being  
employed for computing parallax distances in terms of redshift for various specific world models (von Hoerner 1974).
One can simplify it to some extent by getting rid of the square-root in the denominator (Peacock 1999),
to write it in an alternate form,
\begin{eqnarray}
\label{eq:10}
d_{\rm p}=\frac{c}{H_{\rm o}}\frac{q_{\rm o}z+\left(q_{\rm o}-1\right)\left(-1+\sqrt{1+2q_{\rm o}z}\right)}
{\left(q_{\rm o}-1\right)\left(q_{\rm o}-1-q_{\rm o}z\right)+\left(2q_{\rm o}-1\right)\sqrt{1+2q_{\rm o}z}}.
\end{eqnarray}
The above relations for the parallax distance are inconvenient when to be used for small $q_{\rm o}$ and/or $z$ as then one 
has to evaluate 
these expressions in limit. One can instead of Eq.~(\ref{eq:7}) use an alternate expression for $r$ (Terrell 1977) 
\begin{eqnarray}
\label{eq:11}
r=\frac{c}{H_{\rm o}R_{\rm o}}\frac{z}{(1+z)}\frac{\left[1+z+\sqrt{1+2q_{\rm o}z}\right]}
{\left[1+q_{\rm o}z+\sqrt{1+2q_{\rm o}z}\right]},
\end{eqnarray}
and along with Eqs.~(\ref{eq:4}), (\ref{eq:8}), after some algebraic manipulations, one gets $d_{\rm p}$ as,
\begin{eqnarray}
\label{eq:12}
d_{\rm p}=\frac{c}{H_{\rm o}}\frac{z\left[1+z+\sqrt{1+2q_{\rm o}z}\right]}
{\left[(1+z+z^2+q_{\rm o}z-q_{\rm o}z^2)+(1+z)\sqrt{1+2q_{\rm o}z}\right]}.
\end{eqnarray}

\section{THE CORRECTION}
In the derivation of the expression for the parallax angle, while tracing the light path from the source 
to the observer, the two baseline ends were defined by a set of comoving coordinates (Weinberg 1972). But the two ends of any rod or baseline, 
be it the sun-earth line or some larger baseline in the solar system (or still larger ones but as long as one is confined within 
a gravitationally bound system like our galaxy), cannot be freely falling with the expanding cosmic fluid as the distance 
between the two ends of the rod is taken to be fixed. One can consider one end of the baseline to be at rest with respect to the 
underlying cosmic fluid, but then the other end, at a fixed proper distance $b$, will have a velocity $v=-H_{\rm o} b$ along the baseline, 
with respect to 
the underlying comoving substratum. That means the second end of the baseline, because of its motion with respect to the comoving 
substratum, will register a stellar aberration,  $\delta \approx b H_{\rm o}/c$ toward the baseline direction. The stellar aberration 
will add to the  value of the parallax angle as given by expression~(\ref{eq:3}) and the actually measured parallax angle will be, 
%\pagebreak

\begin{eqnarray}
\label{eq:13}
\Psi=\psi+\delta=b\left[\frac{\left(1-k\,r^2\right)^{1/2}}{R_{\rm o}\,r}+\frac{H_{\rm o}}{c}\right].
\end{eqnarray}
Then writing the parallax distance $D_{\rm p}=b/\Psi$, in place of (\ref{eq:4}) we have a modified value,
\begin{eqnarray}
\label{eq:14}
D_{\rm p}=\frac{R_{\rm o}\,r}{(1-k\,r^2)^{1/2}+R_{\rm o}\,rH_{\rm o}/c}.
\end{eqnarray}
In the $\Omega_\Lambda = 0$ cosmologies, the modified formula for parallax distance then becomes,
\begin{eqnarray}
\label{eq:15}
D_{\rm p}=\frac{c}{H_{\rm o}}\frac{\left[q_{\rm o}z+\left(q_{\rm o}-1\right)\left(-1+\sqrt{1+2q_{\rm o}z}\right)\right]}
{\left[\left(q_{\rm o}-2\right)\left(q_{\rm o}-1-q_{\rm o}z\right)+\left(3q_{\rm o}-2\right)\sqrt{1+2q_{\rm o}z}\right]}.
\end{eqnarray}
Using relation~(\ref{eq:11}), we get an alternate form  for $D_{\rm p}$ as,
\begin{eqnarray}
\label{eq:16}
D_{\rm p}=\frac{c}{H_{\rm o}}\frac{z\left[1+z+\sqrt{1+2q_{\rm o}z}\right]}
{\left[(1+2z+2z^2+q_{\rm o}z-q_{\rm o}z^2)+(1+2z)\sqrt{1+2q_{\rm o}z}\right]}.
\end{eqnarray}

The expression~(\ref{eq:16}) is much simpler to use, especially when evaluating $D_{\rm p}$ for small $q_{\rm o}$ or $z$ values 
as one can avoid going through the process of taking a limit. 
For example, for low redshifts one gets $D_{\rm p}={cz}/{H_{\rm o}}$ from Eq.~(\ref{eq:16}) in a straightforward manner, while it is 
somewhat involved to get the same expression from Eq.~(\ref{eq:15}) that requires a series expansion and the cancellation of various 
terms. Of course to evaluate Eq.~(\ref{eq:15}) for the $q_{\rm o}\rightarrow 0$ case is even more cumbersome, 
while it is quite easily done for Eq.~(\ref{eq:16}). 
\section{DISCUSSION}
The parallax distance has some peculiarities, especially in limiting cases. It was Lobachevsky in 1829 (see North 1965) 
who first brought attention to the curious fact that in a hyperbolic world there is a certain minimum finite value of the parallax 
angle. Strangely it happens so even though the extent of the universe is infinite.  On the other hand 
Schwarzschild (1900) pointed out that in a finite universe with a positive curvature, the parallax distance 
increases and becomes infinite (with a zero parallax angle) at a certain point (even in a finite-sized universe) 
and that for further objects the parallax distance decreases, becoming finite. 
Even in the case of an expanding universe such peculiarities exist. From Eq.~(\ref{eq:16}), for $z \rightarrow \infty$, we get,
\begin{eqnarray}
\label{eq:19}
D_{\rm p} =\frac{c}{H_{\rm o}}\left[\frac{1}{2-q_{\rm o}}\right].
\end{eqnarray}
Thus while parallax distance for a source at the horizon (i.e., $z \rightarrow \infty$) is always finite in a closed universe 
for $0.5<q_{\rm o}<2$, it could become infinite for $q_{\rm o}=2$ case (even though the universe is of finite extent). 
However, for still higher $q_{\rm o}$ values the parallax distance would first be 
finite at low redshifts, become infinite at some finite redshift (for example, at z=9.1 for the $q_{\rm o}=5$ case) and again become 
finite for yet higher redshifts till one reaches the horizon. The negative sign for $D_{\rm p}$ in such cases simply reflects 
the fact that the corresponding parallax angle would be negative, like the diverging directions of the north pole observed 
from two ends of a baseline at a parallel in the southern hemisphere. On the other hand for all open (infinite extent) 
universe models ($q_{\rm o}\leq 0.5$), even the most distant observable source (at the object horizon!) has a maximum 
finite parallax distance ($D_{\rm p} \leq 2c/3H_{\rm o}$) and thus a minimum finite parallax angle. This needs to be carefully taken 
into account when using distant objects as the background field against which the parallax of a foreground object is to be measured. 
%\pagebreak
\begin{table*}[t]
%\begin{table}[t]
\caption{Parallax for various FRW ($\Lambda=0$) world models}
\begin{tabular}{@{}ccccccccccc}
%\begin{tabular}{@{\hspace{4mm}}c@{\hspace{4mm}}c@{\hspace{4mm}}c@{\hspace{4mm}}c@{\hspace{4mm}}c@{\hspace{4mm}}c}\\
\hline
$k$ && $q_{\rm o}$ && World model & $\left(\frac{H_{\rm o}}{c}\right)d_{\rm p}$  
&&$\left(\frac{H_{\rm o}}{c}\right)D_{\rm p}$ &$\psi_{z \rightarrow \infty}$ &$\Psi_{z \rightarrow \infty}$  \\
\hline
\hline
+1 && 1 && spherical & $\frac{z}{\sqrt{1+2z}}$ && $\frac{z}{z+\sqrt{1+2z}}$ & $0$ & $\frac{bH_o}{c}$\\
0 && 0.5 && flat & $2\left[1-\frac{1}{\sqrt{1+z}}\right]$ && $\frac{\sqrt{1+z}-1}{(3/2)\sqrt{1+z}-1}$ & $\frac{bH_o}{2c}$ & $\frac{3bH_o}{2c}$\\
-1 && 0 && hyperbolic & $\frac{z(1+z/2)}{{1+z+z^2/2}}$ && $\frac{1}{2}\left[1-\frac{1}{(1+z)^2}\right]$ & $\frac{bH_o}{c}$ & $\frac{2bH_o}{c}$\\
\hline
\end{tabular}
\end{table*}
%\end{table}

\begin{figure}
%\epsscale{0.6}
%\plotone{fig1.eps}
\scalebox{0.5}{\includegraphics{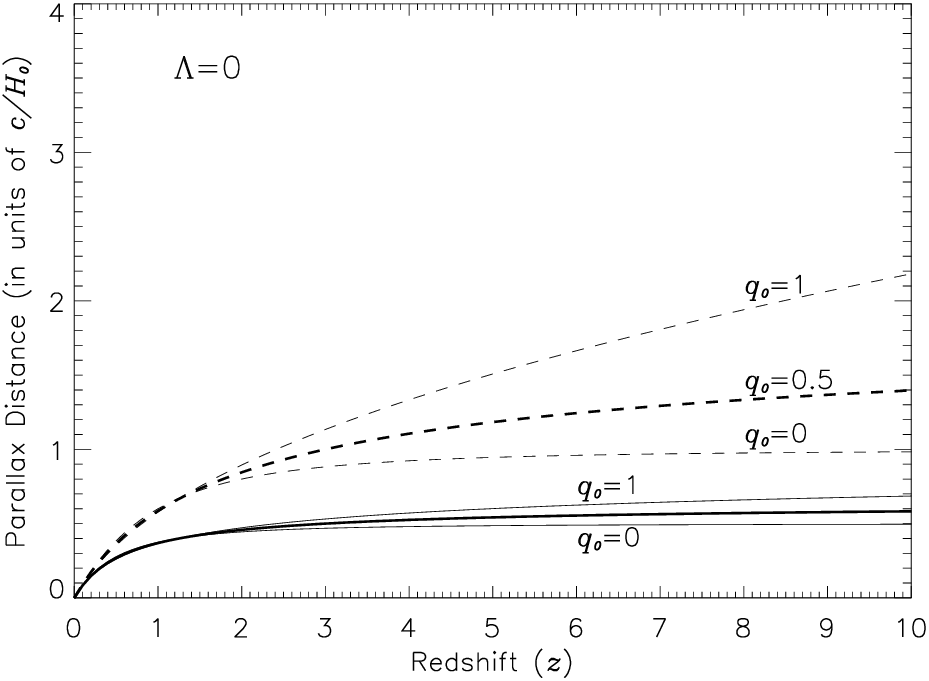}}
\caption{A plot of the parallax distance with redshift for different world models in $\Lambda=0$ 
cosmologies. The upper three curves (broken lines) are for $d_{\rm p}$, calculated from the existing expressions  
in the literature, while the lower plots (continuous lines) for the same three world models are for $D_{\rm p}$, 
calculated using the modified formulae. The bold lines are for the flat-space world model ($q_{\rm o}=0.5$).}
\end{figure}

As both the parallax angle and the correction for the parallax angle are proportional to the baseline $b$, the 
relative correction 
to the parallax is independent of the length of the baseline and as we mentioned earlier, the correction could 
become appreciable at 
large redshifts. From the expression~(\ref{eq:12})  for $d_{\rm p}$ we find that as $z  \rightarrow \infty$, $d_{\rm p}$ 
reduces to $c/[H_{\rm o}(1-q_{\rm o})]$ while $D_{\rm p}$ from Eq.~(\ref{eq:16}) becomes $c/[H_{\rm o}(2-q_{\rm o})]$. 
Thus the corrected parallax-distance values are smaller by a factor $(2-q_{\rm o})/(1-q_{\rm o})$ at large redshifts, 
which for $q_{\rm o}= 0$ case (Milne's world model) 
is a reduction factor of 2 and in the flat space ($q_{\rm o}= 0.5$) it is a factor of 3, the factor could be larger for 
positive curvature spaces with $q_{\rm o}>0.5$.
In Table 1 we have shown a comparison of the formulae for older incorrect parallax distance $d_{\rm p}$ 
(c.f. von Hoerner 1974) and the corrected parallax distance $D_{\rm p}$ for some simple representative FRW world models 
($q_{\rm o}= 0,\; 0.5,\; 1$) in the $\Omega_\Lambda = 0$ cosmologies. Also tabulated are the minimum values  
of the parallax angles, $\psi$ and $\Psi$ at the horizon ($z \rightarrow \infty$) for each model.
It is interesting to note that for Milne's universe ($q_{\rm o}= 0$) parallax 
distance is equal to the angular diameter distance, but the two differ in all other world models (contrary to 
what mentioned by Ding \& Croft 2009). Figure (1) shows a plot of the $d_{\rm p}$ as well as $D_{\rm p}$, for the same three 
world models. We see that at large redshifts the corrected values for the parallax distance are at least a 
factor of two or three lower depending upon $q_{\rm o}$. Also the parallax distance $D_{\rm p}$ does not seem to increase 
indefinitely with redshift. 

Recent observations have indicated that $\Omega_\Lambda \neq 0$ and that the space may be flat with $k=0$ 
(Spergel et al 2007; Ade et al. 2014). In such a case, $D_{\rm p}$ has to be evaluated numerically from,
\begin{eqnarray}
\label{eq:17}
D_{\rm p}=\frac{c}{H_{\rm o}}\; \frac{\cal I}{1+\cal I},
\end{eqnarray}

\begin{figure}
%\epsscale{0.6}
%\plotone{fig2.eps}
\scalebox{0.48}{\includegraphics{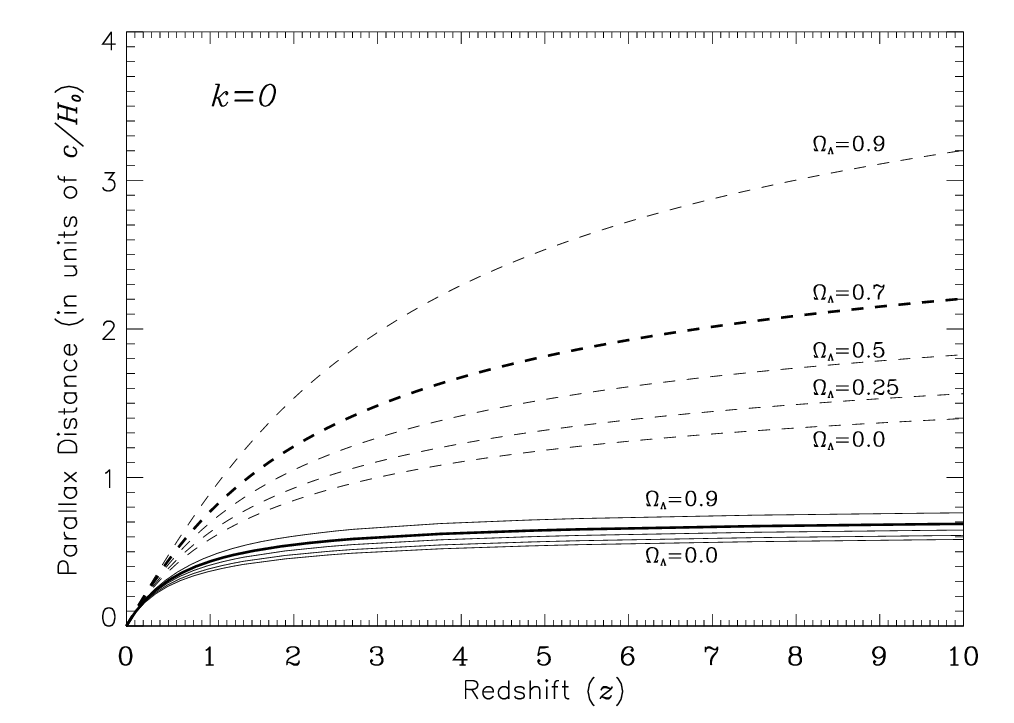}}
\caption{A plot of the parallax distance with redshift for different world models in $k=0$ (flat 
space) cosmologies. The upper five curves (broken lines) are for $d_{\rm p}$ calculated from the existing expressions 
in the literature for different $\Omega_\Lambda$ values, while the lower plots (continuous lines) are for the same 
five world models for $D_{\rm p}$, calculated using the modified formulae. The bold lines are for the presently 
most-favored model $\Omega_\Lambda=0.7$.}
\end{figure}
where 
\begin{eqnarray}
\label{eq:18}
{\cal I}=\int^{1+z}_{1}\frac{{\rm d}z}{\left(\Omega_\Lambda+\Omega_mz^3\right)^{1/2}}.
\end{eqnarray}

Figure (2) shows a plot of parallax distance 
with redshift for the flat space for different $\Omega_\Lambda$ values, including the most likely 
value $\Omega_\Lambda=0.7$ as 
inferred from the Wilkinson Microwave Anisotropy Probe (WMAP) as well as the Planck observations 
(Spergel et al 2007; Ade et al. 2014). We see that at large redshifts, the parallax distances calculated 
from the modified expressions could be lower than from the older incorrect ones by as much as a factor of three or more  
in the currently favored cosmologies (viz. $\Omega_\Lambda=0.7, k=0$).

There are a number of proposals in the literature for employing parallax measurements for cosmological purposes.
(Elvis \& Karovska 2002; Jain \& Ralston 2008; Quercellinin et al. 2009; Ding \& Croft 2009). 
For example, it has been suggested (Kardash\"ev 1986; Ding \& Croft 2009) that for all distant extragalactic objects, like e.g. quasars, the peculiar 
motion of the solar system would result in a parallax  which will be increasing with 
time. Over, say, a ten year period a velocity of 369 km s$^{-1}$ (as inferred from WMAP observations of the CMBR by 
Hinshaw et al. 2009) will give 
rise to a baseline of $\sim 10^{16}$ cm, from which parallax measurements could be made of distant objects. Then one does not 
have to wait for attainment of nano-arcsec resolutions. Such a large baseline,   
even with $\mu$-arcsec ($<10^{-11}$ radians) resolutions, could take us to cosmological 
distances ($>10^{27}$ cm). However, we should point out here that contrary to the view expressed by Ding \& Croft (2009), 
the parallax in such a measurement will be given 
by the corrected expression for $D_{\rm p}$ (Eq.~(\ref{eq:16})) and not by $d_{\rm p}$ (Eq.~(\ref{eq:12})), as would have been 
the case if one were using two independent points comoving with
the expanding cosmic substratum as baseline ends (see also R\"as\"anen 2014). In this case, it is the same observer (at Earth) shifted by a distance 
due to its peculiar motion and not really freely falling with the cosmic fluid at the new position. Let us say, the observer `O', 
due to its 
peculiar velocity $V$, moves from point `A' to `B' in the cosmic substratum, and that `B' has a motion $V_{\rm o}$ with 
respect to `A' due to the cosmic expansion. Assuming no other gravitation interactions affecting the motion of `O', it will 
continue to have $V$ with respect to `A' and $V-V_{\rm o}$ with respect to `B'. Thus `O' will have a parallax 
$D_{\rm p}$ with the baseline of a fixed length `AB', any stellar aberration due to its motion  $V$ will be the same at 
either end of this fixed baseline, and thus will not affect the parallax measurements. 
However. it should be also noted that even a small change in Earth's velocity  $V$ by say, a few cm s$^{-1}$, 
between the two epochs of the observations, could give rise to a stellar aberration in tens of $\mu$-arcsecs which 
would mimic the parallax measurements.
On the positive side, we may add that a recent value of the solar motion (Singal 2011; Rubart \& Schwarz 2013; Tiwari et al. 2014) 
measured with respect to a reference frame of 
faint radio sources from the NRAO VLA Sky Survey (NVSS, Condon et al. 1998), which contains 1.8 million sources 
with a flux-density limit $S>3$ mJy at 1.4 GHz, is found to be about four times larger than the CMBR 
value. This rather high value would imply even much bigger baselines for 
the parallax measurements and in fact the parallax measurements of distant quasars could in turn help resolve the discrepancy of solar 
motion with respect to the two different reference frames, namely CMBR and NVSS.

At the same time while making parallax measurements, one has to be wary of some other, normally unexpected, sources of errors 
(See also Novikov 1977). Any light ray that passes close to a sufficiently massive body, undergoes a gravitational bending. 
For example, a spherical mass $M$ causes a gravitational bending by an angle $4GM/Rc^2$ for a light ray passing at an impact 
parameter $R$ (Weinberg 1972). Thus a ray grazing Sun's disc gets 
deflected by 1.75 arcsec. For a baseline comprising two opposite ends on Earth's orbit around Sun, light from a distant source, 
even $90^\circ$ away in sky from the ecliptic plane, could have a relative deflection which could be as much as $\sim 8$ milli-arcsec due to Sun's 
gravity. A light ray from the same source reaching Earth could get deflected even by Jupiter's mass (when Earth and Jupiter are at their 
closest distance) by an angle about a $\mu$-arcsec. 
The gravitational deflections could be much larger for the sources lying closer to the ecliptic plane in their angular positions in the sky.
Moreover the masses and the impact parameters of different solar system bodies encountered by two rays, while en route to two ends 
of the baseline being used for the parallax measurements, could vary. 
A satellite like GAIA, which is aimed to 
determine parallax with an accuracy of 200 $\mu$-arcsec for half a million distant quasars, while observing from two 
diametrically opposite grazing positions around Earth, could have parallax of a distance quasar affected by the differential 
gravitational bending by the Earth up to $\sim 1$ milli arcsec, which is order of magnitude more than the projected accuracy 
(200 $\mu$-arcsec). As GAIA may monitor each of its targets about 70 times during its operation over five years, these 
repeated observations of the same object will have to be carefully corrected for the gravitational bending effects for  
each run separately, in order to achieve the intended accuracy of microarcsecs.
It will be even more crucial for the parallax measurements of 1 billion stars,  
aimed by GAIA to be determined with an accuracy of about 20 $\mu$-arcsecs at 15 mag, and 200 $\mu$-arcsecs at 20 mag, as  
the gravitational bending due to Sun and the planets will affect these measurements as well. In fact 
at the level of these accuracies even higher order terms up to the quadrupole components of 
the gravitational fields of the massive planets like Jupiter and Saturn (see e.g., Kopeikina1 \& Makarov 2007)  could become 
important to be taken into account.

However, it can be hoped that the exact path of the light ray and the relative positions of the solar system bodies could be 
used to calculate their gravitational bending effects for each ray and the true parallax could be extracted from the observed values.
One saving grace could be 
that if a nearby (in angular sky coordinates) background object is being used as a reference point, then most of the above deflections
due to aberration or gravitational bending would be common for the reference point as well as the target source and hence the 
parallax ascertained from relative measurements of the two may not be affected, at least to a first order. However one has to be 
careful if the second order terms could become sufficiently large to be comparable to the accuracy we might seek to achieve.

\end{document}